# Evaluation of an AI System for the Detection of Diabetic Retinopathy from Images Captured with a Handheld Portable Fundus Camera: the MAILOR AI study

T W Rogers, PhD[1,*]; J Gonzalez-Bueno, MRes[1]; R Garcia Franco, MD[2]; E Lopez Star, MD[2]; D Méndez Marín, MD[2]; J Vassallo, MD FICO FEBO[3]; V C Lansingh, MD, PhD[2]; S Trikha, MBA FRCOphth[1,4]; N Jaccard, PhD[1].

**Affiliations**

[1]Visulytix Ltd, Screenworks, Highbury East, London, N5 2EF, United Kingdom

[2]Mexico, IMO. Instituto Mexicano de Oftalmología. INDEREB. Instituto de la Retina del Bajío

[3]Mater Dei Hospital, Triq Dun Karm, Imsida, Malta

[4]King's College Hospital NHS Foundation Trust, London, SE5 9RS, United Kingdom

*Corresponding author: Thomas W. Rogers, tom@visulytix.com

**Word Counts:**

Abstract: 250

Article: 3254

# Abstract

**Objectives**: To evaluate the performance of an Artificial Intelligence (AI) system (Pegasus, Visulytix Ltd., UK), at the detection of Diabetic Retinopathy (DR) from images captured by a *handheld* portable fundus camera.

**Methods**: A cohort of 6,404 patients (~80% with diabetes mellitus) was screened for retinal diseases using a handheld portable fundus camera (Pictor Plus, Volk Optical Inc., USA) at the Mexican Advanced Imaging Laboratory for Ocular Research. The images were graded for DR by specialists according to the Scottish DR grading scheme. The performance of the AI system was evaluated, retrospectively, in assessing Referable DR (RDR) and Proliferative DR (PDR) and compared to the performance on a publicly available desktop camera benchmark dataset.

**Results**: For RDR detection, Pegasus performed with an 89.4% (95% CI: 88.0-90.7) Area Under the Receiver Operating Characteristic (AUROC) curve for the MAILOR cohort, compared to an AUROC of 98.5% (95% CI: 97.8-99.2) on the benchmark dataset. This difference was statistically significant. Moreover, no statistically significant difference was found in performance for PDR detection with Pegasus achieving an AUROC of 94.3% (95% CI: 91.0-96.9) on the MAILOR cohort and 92.2% (95% CI: 89.4-94.8) on the benchmark dataset.

**Conclusions**: Pegasus showed good transferability for the detection of PDR from a curated desktop fundus camera dataset to real-world clinical practice with a handheld portable fundus camera. However, there was a substantial, and statistically significant, decrease in the diagnostic performance for RDR when using the handheld device.

Introduction

Diabetic Retinopathy (DR) imparts a significant burden on healthcare services worldwide. It is estimated that one third of the 285 million people with diabetes worldwide will have some degree of DR and 50% of patients with proliferative diabetic retinopathy will lose vision within 5 years.[1] By 2030, the number of people with diabetes is likely to reach 552 million.[1] In contrast, the growth in the number of ophthalmologists is unlikely to match the increasing demand.[2,3]

Deep learning-based Artificial Intelligence (AI) systems have shown promise across a number of healthcare imaging domains. For example, Esteva et al. describe a study whereby a group at Stanford University trained a system to classify skin lesions from photographs.[4] The performance of detecting malignant melanomas and carcinomas was found to match that of 21 board-certified dermatologists. Similar networks have been developed to detect breast cancer with equivalent accuracy to experts.[5]

Within the field of ophthalmology, algorithms to detect and classify DR have been developed since at least 2010.[6] However, to our knowledge, no research is available on the performance of such deep learning-based systems with portable *handheld* cameras, which are often used in real-world clinical screening programs. There is little evidence that the performance of deep learning systems, trained and tested on conventional desktop cameras, is transferable to images from handheld devices.

In this study, we evaluate the performance of a deep learning system (Pegasus, Visulytix Ltd., UK) at detecting Referable DR (RDR) and Proliferative DR (PDR), retrospectively, on a cohort of patients screened at the Mexican Advanced Imaging Laboratory for Ocular Research (MAILOR). These results are compared to a "benchmark" performance of the software as determined on a publicly available dataset of curated fundus images from a conventional desktop device.[7]

## Materials and Methods

### Data Collection

The data was collected at the Mexican Advanced Imaging Laboratory for Ocular Research (MAILOR). This study adhered to the tenets of the Declaration of Helsinki. A cohort of participants was randomly selected from patients who attended the clinic between October 2016 and March 2017. Patients were screened for retinal diseases using a 40-degree Field-of-View (FoV) non-mydriatic handheld fundus camera (Pictor Plus, Volk Optical Inc., USA). Images were analysed and graded by a board of experts. Each grader was certified from both the Singapore Eye Research Institute and Melbourne University for fundus image reading.

The Indian Diabetic Retinopathy image Dataset (IDRiD),[7] an independent and publicly available dataset, was employed as a "benchmark" dataset. This dataset is highly curated and contains good quality images captured from a conventional mydriatic 50-degree FoV desktop camera (VX-10 alpha, Kowa American Corp., USA). The images have a resolution of

4288-by-2848 pixels and were captured by a retinal specialist at an eye clinic located in Nanded, Maharashtra, India. The dataset is representative of the Indian population and there was no overlap between patients enrolled in the MAILOR study, and patients screened in the benchmark dataset. The dataset is not necessarily representative of clinical practice, but we believe that it provides a good benchmark for the upper bound on the range of diagnostic performance of the software.

Handheld cameras offer some benefits compared to desktop ones when used in screening programs. In particular, they offer mobility and flexibility, making it easier to screen patients in remote locations. However, the quality of the resultant images is usually lower than conventional desktop cameras.[8] Figure 1 shows examples from the MAILOR study and the benchmark dataset, and highlights the difference in image quality.

## Screening Protocol

In the MAILOR study, four fundus images were captured per patient; one disc-centred and one macula-centred per eye. The quality of each image was assessed to determine whether it was adequate for the purposes of grading. This includes the following rules: (i) an image with visible referable retinopathy or maculopathy is always adequate for grading; and (ii) an image with adequate field definition (entire optic disc imaged and a fovea at least two disc diameters from the image edge) and clarity (third-generation vessels radiating around the fovea must be visible) is adequate for grading.[9]

Images identified as ungradable by the graders were not assessed for retinal disease. Each eye with at least one gradable image was assessed for DR and graded by at least one grader, providing DR and Diabetic Maculopathy (DM) grades according to the Scottish grading system.[9] Second readings were performed in approximately 10% of examinations for quality assurance. In these cases, if a discrepancy with the first reader was identified, the image was graded again by a third expert who took into account both the image and the grades of the initial graders. This was taken as the final grade.

In total, 25,616 fundus images were captured from the initial MAILOR cohort of 6,404 patients. Images from 5,752 patients were subsequently analysed, after excluding unlabelled and ungradable images (Figure 2). Patient and image characteristics are shown in Table 1. It was found that the prevalence of diabetes mellitus in the MAILOR cohort (~80%) was higher than the expected population prevalence. This could be attributed to the fact that the participants are in most cases newly diagnosed diabetic patients. A lower than expected DR prevalence was found in the cohort, given the number of diabetic patients. This could be explained by the fact that patients already diagnosed with diabetic retinopathy might not consider it necessary to enter the study.

The benchmark dataset (IDRiD) consists of 516 fundus photographs graded by experts for diabetic retinopathy according to the International Clinical Diabetic Retinopathy (ICDR) disease severity scale.[10] All benchmark images are macula-centred, with a clear view of the optic disc. Experts also verified that all images were of adequate quality, clinically relevant, that no image was duplicated, and representative of all DR grades.[7]

## Data Processing

In this study, the images were uploaded to the deep learning system, Pegasus (Visulytix Ltd., UK), and the performance of its DR detection feature was assessed. Pegasus is a decision support tool for screening for a number of major eye diseases. For DR screening, Pegasus provides a DR grade per image, as well as visual representations of detected microaneurysms, exudates and haemorrhages overlaid on the original image. The system also provides a quality control feature to confirm, for each image, that the correct type of image was uploaded, and that it is of sufficient quality to be graded. The DR output from Pegasus was used to generate the prediction for the presence of RDR and/or PDR for each patient.

Pegasus assesses images for diabetic retinopathy according to the ICDR scale[10] in order to integrate with the majority of clinical workflows used around the world. However, MAILOR employs the Scottish grading scheme.[9] RDR corresponds to a DR of grade 2 or above in the ICDR scale, and in the Scottish scheme, RDR is defined as R2 or above, or DM 1. In both systems, PDR is defined by the same characteristics.

The use of two different grading schemes can lead to disagreements when grading an image for RDR. This occurs when haemorrhages are present in the retina, but none of the hemifields contains four or more of them, or when there are exudates present in the image farther than one disc diameter from the fovea. In these cases, the image would be considered as referable under the ICDR scale, but not by the Scottish scheme.

All images were uploaded to a locally hosted version of the Pegasus platform using a Python-based Application Programming Interface (API) in a continuous fashion with 8 image processing requests made in parallel from the client computer. No image pre-processing was applied prior to upload; the raw images were uploaded directly to Pegasus.

For the MAILOR study, diagnostic performance was assessed in three ways: (i) on a patient-by-patient basis where the diagnosis for the patients was taken as the worst outcome from Pegasus across the four images acquired; (ii) on an image-by-image basis considering disc-centered fields only; (iii) on an image-by-image basis considering macula-centered fields only.

## Statistical Analysis

The performance of the AI system was measured for RDR and PDR. The prediction for the presence of RDR and/or PDR for each patient was compared to the reference standard. The performance was assessed using Receiver Operating Characteristic (ROC) analysis to determine the sensitivity, specificity and Area Under the ROC curve (AUROC). Confidence Intervals (CI) at the 95% significance level were determined using bootstrap sampling with 1,000 replications for all metrics reported. Randomised permutation testing with 1,000 samples was used for testing whether differences in performance were statistically significant. Statistical significance was defined at $p<0.05$.

## Results

There were no processing errors for any of the images. For MAILOR, processing was performed at an average speed of 2.16 seconds/image for the 22,180 images. The IDRiD dataset was processed in a similar way but with an average speed of 1.07 seconds/image for 516 images. Pegasus gave average gradability scores of 84.8% (standard deviation: 19.6%) and 44.6% (standard deviation: 35.1%) for IDRiD and MAILOR images, respectively.

Results for Pegasus for both the MAILOR study and the benchmark dataset are shown in Table 2. On images from the handheld camera, Pegasus achieved an AUROC of 89.4% (95% CI: 88.0-90.7) when detecting RDR, corresponding to a sensitivity of 81.6% (95% CI: 79.0-84.2) and at a specificity of 81.7% (95% CI: 80.9-82.6), at an operating point chosen such that the sensitivity and specificity were approximately equal. When detecting PDR, Pegasus obtained an AUROC of 94.3% (95% CI: 91.0-96.9), corresponding to a sensitivity of 86.6% (95% CI: 78.9-93.5) and specificity of 87.7% (95% CI: 87.0-88.4). The corresponding ROC curves are shown in Figure 3. Selected examples of false positives are given in Figure 5.

When assessed on the desktop camera benchmark dataset, Pegasus obtained an AUROC of 98.5% (95% CI: 97.8-99.2) for RDR, with corresponding sensitivity and specificity of 93.4% (95% CI: 90.8-95.8) and 94.2% (95% CI: 91.0-97.2), respectively. For PDR, the AI system achieved an AUROC of 92.2% (95% CI: 89.4-94.8), with a sensitivity of 83.7% (95% CI: 74.5-91.8) and specificity of 84.6% (95% CI: 81.5-87.7). The corresponding ROC curves are shown in Figure 6.

The AUROC difference for RDR detection was substantially lower for patients from the MAILOR study than the IDRiD benchmark dataset. This difference was statistically significant (p<0.001). There was approximately a 12-13% disparity in the RDR sensitivity/specificity performances between the handheld and benchmark desktop devices. For PDR, Pegasus performed slightly better on the MAILOR cohort than the benchmark dataset in terms of AUROC, however this difference was not statistically significant (p=0.172).

Additionally, the RDR performance of Pegasus was compared when using only macula-centred images for a patient against disc-centred images. It was determined that there was a statistically significant difference in AUROCs (p=0.003). The RDR AUROC performance improved by 2.3% when using a macula-centred field over a disc-centred field.

## Discussion

The performance of Pegasus for RDR was statistically significantly, and substantially, lower on the MAILOR cohort using the handheld portable camera than on the benchmark dataset using the conventional desktop camera. There are a number of possible factors that could explain this:

1. The Pegasus software performance for RDR does not generalise well from a curated conventional desktop dataset to a real-world patient cohort imaged using a handheld portable fundus camera due to poor image quality and other factors;

2. The mismatch in the grading system expected by the software and the grading system used in the Clinical Reference Standard (CRS), led to an increase in "false positives";

3. The CRS was substantially worse in the MAILOR study than in the curated benchmark dataset, due to the difficulty in assessing subtle DR features such as dot haemorrhages in the poorer quality images from the handheld device;

4. Inclusion of disc-centred images leads to substantially worse diagnoses for a patient over just scanning their macula.

The answer is likely a combination of the above factors, however, it appears that the biggest contributing factor is the mismatch in grading systems. This is because there is evidence that the software algorithm does generalise in the case of PDR where the grading systems perfectly coincide, and it was observed that there were a number of RDR false positives (see Figure 4) which were due to grading system mismatches. In addition, the performance on the disc-centred field is statistically significantly worse than on macula-centred fields, although not substantially so. This implies the effect of disc-centred images on overall patient diagnosis is small. Image quality may play a role since the Pegasus quality control feature detected that the quality of the images from the handheld camera is substantially lower than for the desktop camera, which can have an effect on the detection of DR.

Overall Pegasus performed well for PDR detection on both the handheld portable camera (MAILOR cohort) and the curated benchmark dataset (IDRiD), with no statistically significant difference between the two image types. This implies that the software algorithm can generalise to other image types, at least in the case of PDR.

The main limitations of this study include: (i) the CRS was determined by manual grading of images and not the final clinical diagnosis, and (ii) the mismatch between the CRS grading system and the software grading system. While these limitations complicate the interpretation of the results for RDR, in particular, they raise some important points about the use of deep learning software in clinical practice. First, organisations should ensure that the deep learning system uses the same grading system as the rest of the screening program, to prevent an apparent increase in false positive referrals. Second, deep learning software such as Pegasus, should clearly identify the types of imaging devices that give optimal performance and preferably contain an image quality module which indicates the suitability of images for automated analysis. Finally, caution should be taken when assessing the performance of deep learning software on high quality, curated, public datasets that may not necessarily correspond well to the realities of clinical practise.

Despite these limitations, the performance of Pegasus is favourable when compared to similar systems from studies in the literature. For example, Rajalakshmi et al. determined that a deep learning system obtained a sensitivity of 99.3% (95% CI: 96.1-99.9) and specificity of 68.8% (95% CI: 61.5-76.2) for the detection of RDR from a non-handheld, but portable, smartphone-based fundus camera.[11] Pegasus achieved a sensitivity of 81.6% (95% CI: 83.9-90.2) and a specificity of 81.7% (95% CI: 85.7-89.7) for RDR, which amounts to 17.7% drop in sensitivity but a 12.9% improvement in specificity over the system evaluated by Rajalakshmi et al.[11] In terms of RDR prediction, Rajalakshmi et al.[11] report a sensitivity of 78.1% (95% CI: 63.8- 83.3) and specificity was 89.8% (86.1, 93.4), compared to the 86.6% (95% CI: 78.9-93.5) sensitivity and 87.7% (95% CI: 87.0-88.4) specificity obtained by Pegasus.

The performance of Pegasus, even when a handheld device is used, also compares favourably to the performance reported in the literature using conventional cameras in a clinical setting. In a study by Tufail et al. the sensitivity for RDR detection varied from 85.0% to 93.8%, and the specificity from 19.9% to 52.3%, depending on the deep learning system used.[12] One of the systems evaluated could not handle disc-centred images, and gave a sensitivity of 100% but a specificity of 0%.[10] In this study, it was determined that Pegasus could handle disc-centred images but there was a small decrease in performance, as would be expected since DR features at the periphery would not be captured in disc-centred images. In addition, a paper by Ting et al. showed a sensitivity of between 91 and 100% and a specificity of 91 to 92% for detecting RDR in a large sample of 76,370 images from multi-ethnic cohorts, which is similar to the performance of Pegasus on the benchmark dataset.[13]

In a study by Gulshan et al., a deep learning algorithm was developed and validated for RDR against two datasets of images captured from conventional cameras; EyePACS-1 and Messidor-2.[14] Under the high sensitivity operating point the algorithm obtained a 97.5% sensitivity and 93.4% specificity on EyePACS-1 and 96.1% sensitivity and 93.9% specificity on Messidor-2. Importantly, poor quality images had been removed by clinicians, and so the results reported are not necessarily comparable to the real-world clinical setting and use of a product such as Pegasus as reported in this study or by Tufail et al.[10] Nevertheless, on the conventional camera used in this study as a benchmark, Pegasus performs similarly with 93.4% (95% CI: 90.8-95.8) sensitivity and 94.2% (95% CI: 91.0-97.2) specificity. However, the results could not be compared for PDR, since they are not reported by the authors.[13]

A strength of this study is the evaluation of the deep learning system 'out of the box'. The system was evaluated on unseen, real-world clinical data from a handheld fundus camera. The reported performance of the system, which was not specifically designed for images from handheld cameras, would likely improve if the system had 'learned' from images captured by either the handheld or conventional fundus cameras evaluated. This study was performed on a large patient cohort (5,752 patients; 22,180 images) which is highly representative of the sorts of images that would be acquired in remote retinal screening programmes. Furthermore, the system demonstrates its scalability through parallel analysis of multiple images at once.

## Conclusion

The combination of handheld fundus imaging and AI assistance has great promise in improving the ability to screen patients in remote geographic locations where there is a shortage of ophthalmologists. To our knowledge, this study is the first to evaluate the performance of an artificial intelligence system on fundus images from a fully portable, handheld device, and the first comparison of performance on handheld versus performance on conventional cameras.

Several AI systems for the grading of DR have been released in recent years.[10,12,13] The majority of published work has shown good performance on images captured using desktop fundus cameras. In this study, it was demonstrated that one such AI system, Pegasus, achieves performance comparable to state-of-the-art[11,13,14] on images acquired using a desktop fundus camera[7] yet generalises to images captured by a handheld camera, without adjustments, in the detection of PDR. Due to the limitations of this study, further work will have to be conducted to determine the level of generalisation of RDR.

Artificial Intelligence systems such as Pegasus offer the intriguing possibility of distributed telemedicine, when used in conjunction with handheld screening devices, for accessible and high-quality care. Importantly, one of the benefits of such systems is the lack of variability and speed in decision making. With the number of patients with diabetic retinopathy likely to grow exponentially in the next two decades, newer sustainable models of care will need to be embedded and legacy clinical systems upgraded. Whilst showing considerable promise, further work is required to identify the clinical and economic 'downstream' effects of such Artificial Intelligence systems in practice.

## Acknowledgements

The authors acknowledge Visulytix Ltd. for provision of the deep learning system, Pegasus, and thank Dr Kanwal Bhatia for her contribution.

## Conflicts of Interest

TW Rogers, J Gonzalez-Bueno and N Jaccard are employed by, and own stock options in, Visulytix Ltd.

S Trikha owns shares in Visulytix Ltd and receives honoraria from Visulytix Ltd.

J Vassallo, V. C. Lansingh, E. L. Star, D. M. Marìn and R Garcia Franco declare no conflicts of interest.

## Data Availability

The public benchmark dataset (IDRiD) is available on a CC-BY 4.0 license. The data collected in the MAILOR dataset is available at the discretion of MAILOR.

# Funding

Funding was provided by Visulytix Ltd. in the provision of the Pegasus deep learning software and statistical analysis of results.

# Figure legends

**Figure 1.** Examples of images used in this study. *Left*: From the benchmark dataset (IDRiD), captured using a desktop (VX-10 alpha, Kowa American Corp., USA) fundus camera. *Right*: From MAILOR, captured using a portable handheld (Pictor Plus, Volk Optical Inc., USA) fundus camera.

**Figure 2.** CONSORT-style diagram for the MAILOR cohort.

**Figure 3.** Receiver Operating Characteristic (ROC) curves for Pegasus measured on the MAILOR patient cohort for RDR and PDR. The shaded regions indicate the estimated 95% confidence intervals on the ROC curves.

**Figure 4.** Selected examples of false positives for RDR detection from the MAILOR study. Exudate (left) and haemorrhage (right) detections by Pegasus are overlaid on the images as white bounding boxes and pink and blue heatmaps, respectively. Image zooms are inset for features detected by Pegasus that are difficult to see.

**Figure 5.** Receiver Operating Characteristic (ROC) curves for Pegasus measured on the IDRiD patient cohort for RDR and PDR. The shaded regions indicate the estimated 95% confidence intervals on the ROC curves.

# Tables

**Table 1:** Summary characteristics of eligible patients from the MAILOR cohort and the benchmark dataset.

|  | **MAILOR dataset** | **IDRiD dataset** |
|---|---|---|
| Number of patients (images) | 5,752 (22,180) | - (516) |
| Age, years ± SD (min-max) | 60.0 ± 11.2 (8-100) | - |
| Female, No. (% patients) | 3,785 (65.8%) | - |
| Diabetic, No. (% patients) | 4,646 (80.8%) | - (100%) |
| Fields, No. (% images) <br> • Macula-centred <br> • Disc-centred | <br> 11,090 (50.0%) <br> 11,090 (50.0%) | <br> 516 (100%) <br> 0 (0%) |
| Disease severity distribution (CRS) <br> • No diabetic retinopathy <br> • Referable diabetic retinopathy <br> • Proliferative diabetic retinopathy | <br> 5,017 (87.22%) <br> 595 (10.34%) <br> 60 (1.04%) | <br> 168 (32.6%) <br> 323 (52.6%) <br> 62 (12.0%) |

**Table 2:** Performance of the AI system, Pegasus, for the detection of Referable Diabetic Retinopathy (RDR) and Proliferative Diabetic Retinopathy (PDR) from the benchmark desktop and handheld fundus devices.

| Fundus Camera | DR Severity | Sensitivity, % (95% CI) | Specificity, % (95% CI) | AUROC, % (95% CI) |
|---|---|---|---|---|
| Benchmark (IDRiD): Desktop Kowa VX-10 alpha | RDR | 93.4 (90.8-95.8) | 94.2 (91.0-97.2) | 98.5 (97.8-99.2) |
| | PDR | 83.7 (74.5-91.8) | 84.6 (81.5-87.7) | 92.2 (89.4-94.8) |
| MAILOR study: Handheld Volk Pictor Plus | RDR | 81.6 (79.0-84.2) | 81.7 (80.9-82.6) | 89.4 (88.0-90.7) |
| | PDR | 86.6 (78.9-93.5) | 87.7 (87.0-88.4) | 94.3 (91.0-96.9) |

# Figures

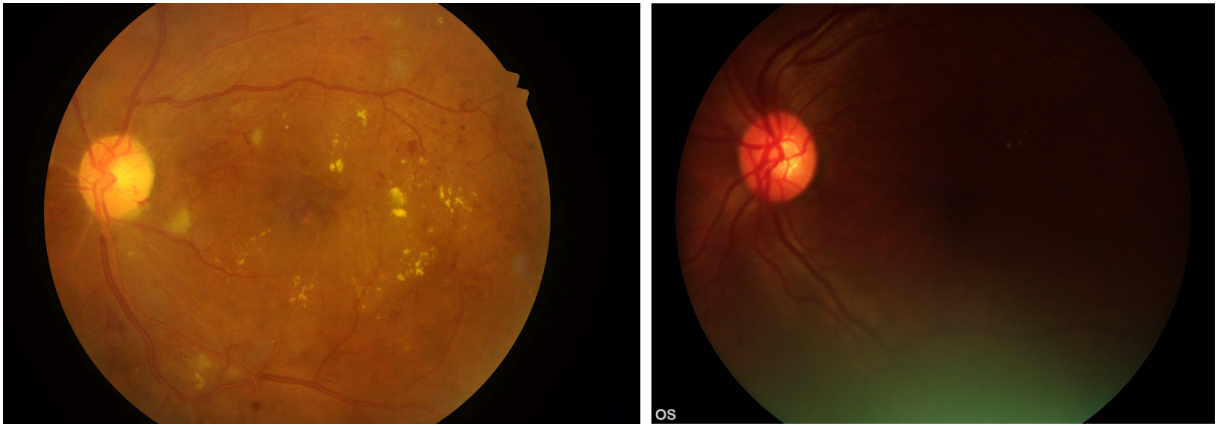

**Figure 1.**

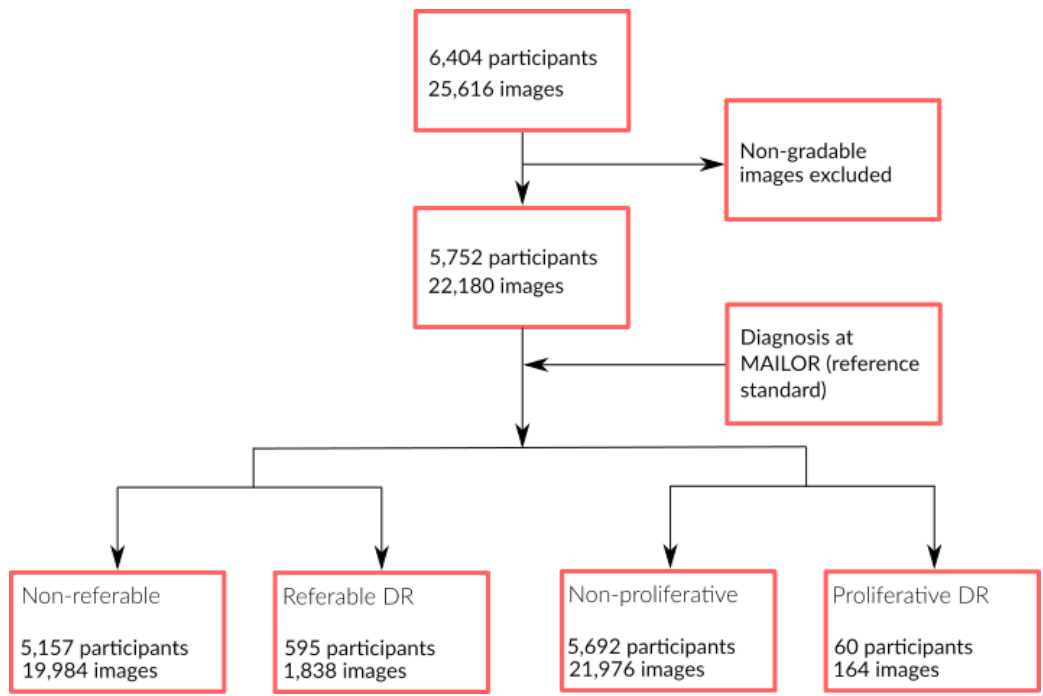

**Figure 2.**

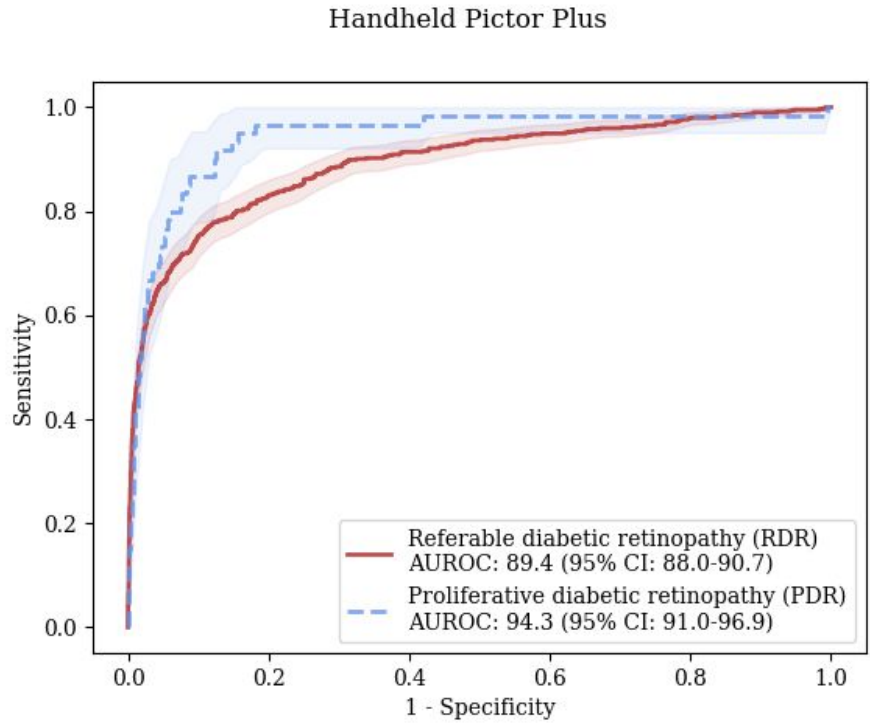

**Figure 3.**

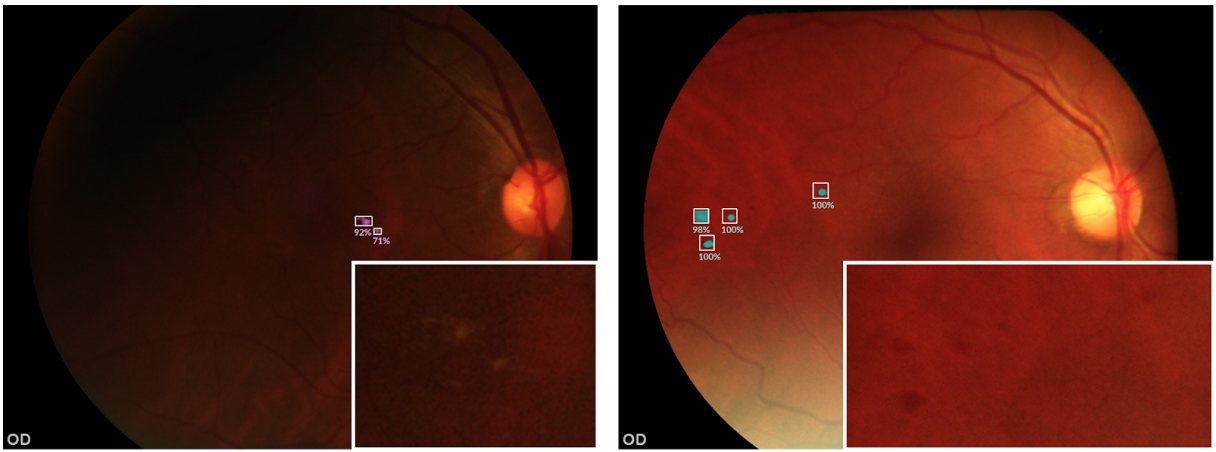

**Figure 4.**

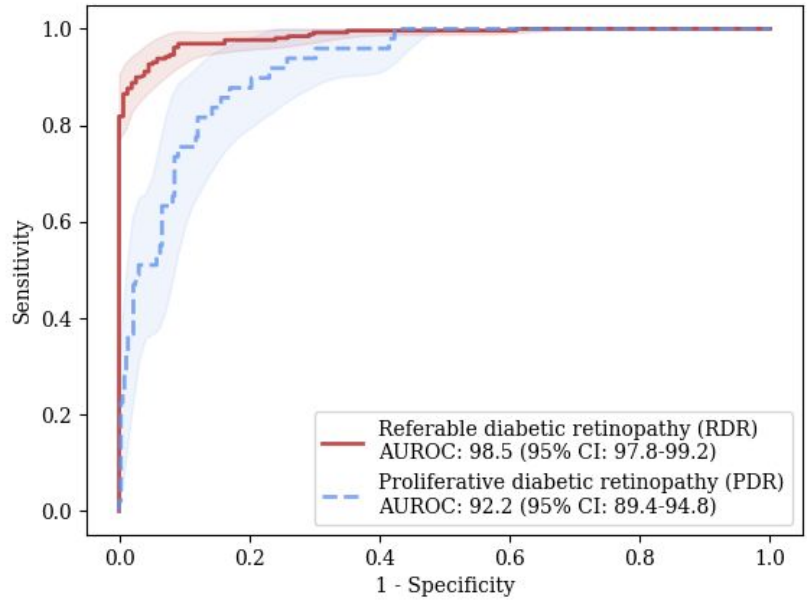

**Figure 5.**